\documentclass[superscriptaddress,twocolumn,preprintnumbers,amsmath,amssymb,floatfix,aps,prb]{revtex4-2}
\usepackage[english]{babel}
\usepackage[colorlinks=true, allcolors=blue]{hyperref}

\usepackage{graphicx,tabularx}
\usepackage[version=3]{mhchem} % Formula subscripts using \ce{}
\usepackage{amssymb}
\usepackage{dcolumn}
\usepackage{color}% to create grayscales for the ball and stick images
\usepackage[mathcal]{euscript}
\usepackage{makecell}
\usepackage{siunitx}

\DeclareSIUnit{\sccm}{sccm}
\DeclareSIUnit{\Torr}{Torr}
\definecolor{gray0}{gray}{0.0}%black
\definecolor{gray64}{gray}{0.25}
\definecolor{gray128}{gray}{0.5}
\definecolor{gray192}{gray}{0.75}
\definecolor{gray255}{gray}{1.0}%white

\newcommand{\review}[1]{#1}

\begin{document}
\title{The impact of strain on the GeV-color center in diamond}
\author{Thijs G.I. van Wijk}
\affiliation{Hasselt University, Institute for Materials Research (imo-imomec), Quantum \& Artificial inTelligence design Of Materials (QuATOMs), Martelarenlaan 42, B-3500 Hasselt, Belgium}
\affiliation{imec, imo-imomec, Wetenschapspark 1, B-3590 Diepenbeek, Belgium}
\author{E. Aylin Melan}
\affiliation{Hasselt University, Institute for Materials Research (imo-imomec), Quantum \& Artificial inTelligence design Of Materials (QuATOMs), Martelarenlaan 42, B-3500 Hasselt, Belgium}
\author{Rani Mary Joy}
\affiliation{imec, imo-imomec, Wetenschapspark 1, B-3590 Diepenbeek, Belgium}
\affiliation{Hasselt University, Institute for Materials Research (imo-imomec), Martelarenlaan 42, B-3500 Hasselt, Belgium}
\author{Emerick Y. Guillaume}
\affiliation{Hasselt University, Institute for Materials Research (imo-imomec), Quantum \& Artificial inTelligence design Of Materials (QuATOMs), Martelarenlaan 42, B-3500 Hasselt, Belgium}
\affiliation{imec, imo-imomec, Wetenschapspark 1, B-3590 Diepenbeek, Belgium}
\affiliation{University of Namur, Namur Institute of Structured Matter (NISM), Rue de Bruxelles 61, 5000 Namur, Belgium}
\author{Paulius Pobedinskas}
\affiliation{imec, imo-imomec, Wetenschapspark 1, B-3590 Diepenbeek, Belgium}
\affiliation{Hasselt University, Institute for Materials Research (imo-imomec), Martelarenlaan 42, B-3500 Hasselt, Belgium}
\author{Ken Haenen}
\affiliation{imec, imo-imomec, Wetenschapspark 1, B-3590 Diepenbeek, Belgium}
\affiliation{Hasselt University, Institute for Materials Research (imo-imomec), Martelarenlaan 42, B-3500 Hasselt, Belgium}
\author{Danny E. P. Vanpoucke}
\email[Corresponding author: ]{Danny.Vanpoucke@UHasselt.be}
\affiliation{Hasselt University, Institute for Materials Research (imo-imomec), Quantum \& Artificial inTelligence design Of Materials (QuATOMs), Martelarenlaan 42, B-3500 Hasselt, Belgium}
\affiliation{imec, imo-imomec, Wetenschapspark 1, B-3590 Diepenbeek, Belgium}

\begin{abstract}% Abstract needs to be <250 words.
Color centers in diamond, such as the GeV center, are promising candidates for quantum-based applications. Here, we investigate the impact of strain on the zero-phonon line (ZPL) position of GeV$^0$. Both hydrostatic and linear strain are modeled using density functional theory for GeV$^0$ concentrations of $1.61$ \% down to $0.10$ \%. We present qualitative and quantitative differences between the two strain types: for hydrostatic tensile and compressive strain, red- and blue-shifted ZPL positions are expected, respectively, with a linear relation between the ZPL shift and the experienced stress. By calculating the ZPL shift for varying GeV$^0$ concentrations, a shift of $0.15$ nm/GPa ($0.38$ meV/GPa) is obtained at experimentally relevant concentrations using a hybrid functional. In contrast, only red-shifted ZPL are found for tensile and compressive linear strain along the $\langle100\rangle$ direction. The calculated ZPL shift exceeds that of hydrostatic strain by at least one order of magnitude, with a significant difference between tensile and compressive strains: $3.2$ and $4.8$ nm/GPa ($8.1$ and $11.7$ meV/GPa), respectively. In addition, a peak broadening is expected due to the lifted degeneracy of the GeV$^0$ $e_g$ state, calculated to be about $6$ meV/GPa. These calculated results are placed in perspective with experimental observations, showing values of ZPL shifts and splittings of comparable magnitude.
\end{abstract}

\maketitle
\section{Introduction}\label{sec:intro}
Over the last few decades, diamond color centers have been seen as promising candidates to fill in the role of single photon emitter for applications such as quantum cryptography\cite{Gisin:2002rmp}, quantum computing\cite{DOHERTY:2013pr, Jelezko:2004prl}, and quantum sensing\cite{hruby:AIP2022}. Color centers arise from defects (which can be missing atoms, impurity atoms, or combinations of both) which modify the optical properties of the crystal by introducing defect levels within the bandgap of the host material. The excitations to and from the defect levels can cause optical transitions, which give rise to a sharp zero-phonon line (ZPL) in the visible spectrum, hence giving color to an otherwise transparent material. Although color centers are not unique to diamond, this material is considered as the superior host for single photon emitters, as its indirect wide bandgap has the capacity to accommodate a wide variety of color centers\cite{Zaitsev:2000prb, Aharonovich:rpp2011}. As of reporting this work, more than 500 diamond color centers have been identified\cite{YELISSEYEV:2003drm}, but only a small fraction have been characterized at the atomic-scale level. Less than 10 of these are considered to be bright and stable enough for a single photon emitter\cite{Aharonovich:rpp2011}.\\
Color centers are often best categorized according to their coherence time, wavelength and Debye-Waller (DW) factor of their ZPL and quantum efficiency. This description makes for a great insight in the possible application of these materials: for instance, quantum communication applications require materials with bright and coherent optical transitions \review{ensuring limited spectral diffusion and}~with long coherence lifetimes\cite{bradac:2019Nc, BersinE:2024APS, bhaskar:2020Nature}.\\
One of the most extensively studied diamond color centers is the Nitrogen-vacancy (NV$^-$) center. This color center has a long coherence time\cite{chen:2020aqt, Beveratos:2001pra}, which would make it a potential material for quantum computing and quantum sensing applications\cite{DOHERTY:2013pr}, if it were not for its weak ZPL emission (DW factor of 0.03 \cite{DOHERTY:2013pr, Riedel:2017prx, zhao:2012JJap}). Other color centers have also been investigated, such as the ones for which the impurity atom belongs to the group IV of the periodic table. \review{They}~are considered to be promising candidates for quantum applications, since they show a bright ZPL emission\cite{RaznkovasL:2021PhysRevB, NeuE:2011NewJPhys, NeuE:2011prb, dietrich:2014njp, palyanov:2015sr, Siyushev:2017prb}. Unlike the NV$^-$ color center, which has a $C_{3v}$ point group, split-vacancy (XV) centers have $D_{3d}$ symmetry, where the atom of element X (Si, Ge, Sn, or Pb) takes the interstitial position between two adjacent vacancies in diamond\cite{hepp:2014prl,GossJP:1996prl,GossJP:2005PRB,GossJP:2007prb,thiering:2018prx}. The resulting inversion symmetry protects the optical transitions from first-order electric field fluctuations, which limits the lifetime-broadening. This is the theoretical limit of the linewidth that can be achieved\cite{RufM:2021JAppPhys, BhaskarM:2017PhysRevLett}.\\
The simplest group IV color center is the negatively charged Silicon-vacancy (SiV$^-$) center. It has a strong ZPL at $738$ nm with a weak sideband (DW factor of $>70$ \%) at room temperature, along with a very short lifetime ($\approx 1$~ns)\cite{NeuE:2011NewJPhys, dietrich:2014njp}. However, it has been shown that it also has a low quantum efficiency and a sub-microsecond spin coherence time\cite{bradac:2019Nc,Rogers:2014prl,Pingault:2014prl}. As a consequence, other group-IV color centers such as the Germanium-vacancy (GeV) center, are receiving increasing interest.\\
The GeV$^-$ center shows similar optical properties as the SiV$^-$ center with regard to linewidth and DW factor, since it has the same symmetry. However, due to its higher mass, GeV$^-$ has a larger ground state splitting, resulting in a longer spin coherence time: In recent experiments, the quantum memory of the negatively charged GeV$^-$ has even been extended to 20 ms\cite{Siyushev:2017prb, SenkallaK:2024PhysRevLett}. In addition, it also shows a higher quantum photoluminescence efficiency compared to SiV$^-$ \cite{TAN:2022ss}, making it an excellent candidate for quantum communication applications.\\
\review{Another path that has been explored to increase the coherence time of split vacancy centers, is by  using the neutrally charged centers instead of the negatively charged ones. The existence of the SiV$^0$ and the GeV$^0$ has been experimentally demonstrated by using electron spin resonance (ESR) measurement\cite{edmondsA:2008prb, dhaenens-johansson:2010prb, komarovskikhA:2018pssa}. The SiV$^0$ and GeV$^0$ are effective spin $S = 1$ systems\cite{edmondsA:2008prb}. Just like their negatively charged counterparts, they exhibit the same D$_{3d}$ symmetry. Neutrally charged split-vacancies possess no partially filled degenerate energy levels in ground state; therefore, these systems are not distorted by the dynamical Jahn-Teller effect. There may however be a product Jahn-Teller effect for the excited state configuration\cite{ThieringG:2019NPJCompMat}. Their spin allows for ESR at room temperature and coherent manipulation of spins. The SiV$^0$ color center (or KUL1 center) has been observed with a ZPL at $946$ nm and a DW factor of 0.90 \cite{RoseB:2018science} It shows insensitivity to environmental decoherence and has a radiative lifetime of $1.8$ ns \cite{RoseB:2018science}. Other neutral group IV color centers are expected to possess similar valuable properties \cite{chen:2020aqt,ThieringG:2019NPJCompMat}. In case of the GeV$^0$ center, a spectral line at $1.979$ eV has been proposed as potential ZPL signature \cite{KrivobokV:2020PRB}, though it was also associated with the vibrational sideband of GeV$^-$ \cite{ makinoY:2024jjap}}.

The present work addresses the characterization and determination of the energy levels of the GeV$^0$ color center, which allows for a finer understanding of the observed effect of strain on the electronic structure of group IV-vacancy color centers, specifically on their ZPL\cite{lindner:2018njp, NeuE:2011prb, grudinkin:2016N,MaryJoy2024pp3873}. Strain is frequently observed as a result of diamond growth, and is especially pronounced in nanocrystalline diamond (NCD). In this paper, we investigate the impact of strain on the GeV$^0$ center electronic structure, more specifically the shift of the GeV$^0$ ZPL line by \textit{ab initio} calculations. Based on these, we propose an interpretation of experimentally observed ZPL shifts for GeV centers, reported earlier by some of the authors.

%% for consistency only Gamma-only PBE ZPL are used in this work
\begin{table}[!tb]
    \caption{Brillouin zone sampling used for the different supercell calculations. The number of atoms in the pristine diamond supercells is given. The color center concentrations are defined as the ratio of GeV centers over C atoms in the system.}\label{table:kpoints}
    \begin{ruledtabular}
    \begin{tabular}{l|rrrrc}
        cell & atoms &  conc.(\%) & PBE relax & PBE \& HSE06 ZPL \\
         \hline
         $c 2\times 2\times 2$ &  $64$ & $1.61$ & $4\times 4\times 4$  & $\Gamma$ \\
         $c 3\times 3\times 3$ & $216$ & $0.47$ & $2\times 2\times 2$  & $\Gamma$ \\
         $c 4\times 4\times 4$ & $512$ & $0.20$ & $2\times 2\times 2$  & $\Gamma$ \\
         $c 5\times 5\times 5$ & $1000$ & $0.10$ & $2\times 2\times 2$ & $\Gamma$ \\
    \end{tabular}
    \end{ruledtabular}
\end{table}

\section{Methodology}\label{sec:meth}
\subsection{Computational Details \label{subsec:compdet}}
Density functional theory (DFT) calculations are performed using the Projector-Augmented-Wave formalism implemented in VASP.5.4.4 \cite{KresseG:PhysRevB1996, KresseG:PhysRevB1999}. The kinetic energy cutoff is set to $600$ eV. The exchange and correlation behavior of the valence electrons (C: $2s^2 2p^2$ and Ge: $3d^{10}4s^2 4p^2$) is described using the generalized gradient approximation as devised by Perdew, Burke, and Ernzerhof (PBE) during atomic structure optimization and initial electronic structure calculations\cite{PBE_PerdewJP:PhysRevLett1996}. Within this work, the standard PBE GW-ready pseudo-potentials are used. Atomic positions and lattice parameters are optimized simultaneously during structure optimization using a conjugate gradient method with the energy convergence criterion set to $1.0\times 10^{-7}$. After relaxation, the maximal forces remaining on any single atom are below $0.53$ meV/\AA. Three structure optimization routes are considered in our work, relevant for the different strain models: (i) only ionic relaxation with the lattice parameter fixed at the PBE diamond bulk lattice parameter ($3.5704$ \AA), indicated as \textit{Bulk}, (ii) ionic relaxation combined with a volume optimization--constrained to a cubic lattice--via a fit to the third order isothermal Birch-Murnaghan EOS, indicated as \textit{Cubic}~\cite{BirchF:PhysRev1947, MurnaghanFD:PNAS1944}, and (iii) a structure optimization allowing ionic positions, cell shape and volume to be optimized simultaneously, indicated as \textit{Skewed}. For the linear and the hydrostatic strain models (\textit{cf.}, Sec.~\ref{sec:model}), the \textit{Bulk} and \textit{Cubic} optimization routes are used, respectively.\\
The first Brillouin zone is sampled using an extended Monkhorst-Pack special $k-$point grid (\textit{cf.}, Table~\ref{table:kpoints}), selected--based on converge tests for the various system sizes--to have the total energy of the system converged to within $1$ meV, and the forces on the atoms within $1$ meV/\AA.
Since local DFT functionals are known to underestimate the bandgap and the position of color center states in the bandgap, we use the corrected range-separated hybrid functional HSE06 for high accuracy electronic structure calculations\cite{HSE03_HeydJ:JChemPhys2003, HSE03Err_HeydJ:JChemPhys2006, HSE06paper_KrukauAV:JChemPhys2006}. However, as the computational cost of HSE06 in the systems investigated here is approximately between one and three orders of magnitude more expensive than PBE, no relaxation are conducted at the HSE06 level and the optimized structures obtained from PBE are used for HSE06 electronic structure calculations\cite{VanpouckeDEP:DiamRelatMater2017}. Since our focus lies with relative changes in electronic structure, the impact of the slight difference in lattice parameter is expected to be negligible.~\review{Furthermore, as DFT is only exact for the ground state of a system, the ZPL is considered as the required absorption energy for excitation, as such to remain within the validity of DFT. This approximation is further validated by the experimental observation that the absorption and emission spectra for the GeV are mirror images\cite{BoldyrevKN:2018JLumen}. In addition, if one were to consider simulating emission, a Franck-Codon description would have to be adopted, which includes a relaxation of the excited state system (which falls outside of the purview of DFT). The associated relaxation energy, however, could be expected to be almost independent of the external strain, as the latter would only be able to slightly modify the hard diamond potential. As a result, its contribution to the relative ZPL position (or ZPL shift due to strain) would be negligible}.\\
Within the field of quantum chemistry, several approaches have been devised to decompose the electron density into atomic charges\cite{BaderRFW:1991ChemRev, ParrRGAyersPWNalewajskiRF:2005JPCA}. In this work, we use Hirshfeld-I charges\cite{BultinckP:2007JCP_HirRef, VanpouckeDannyEP:2013aJComputChem, VanpouckeDannyEP:2013bJComputChem}, which have been shown to be very robust and to correlate well with formal charges\cite{WolffisJJ:MicroMesoMater2019, BosoniE:NatRevPhys2023}. Atomic charges are obtained using the Hirshfeld-I partitioning scheme implemented for grid-based electron densities in the HIVE-code\cite{VanpouckeDannyEP:2013aJComputChem, VanpouckeDannyEP:2013bJComputChem, HIVE_REF}. The convergence criterion for the iterative scheme is set to $1.0\times 10^{-5}$ electron. The charges are integrated on an atom-centered grid with logarithmic spacing in the radial directions (255 points) and spherical Lebedev-Laikov shells of 1202 grid-points\cite{BeckeAD:JChemPhys1987, LebedevVI_grid:1999DokladyMath}.
\begin{figure}[!t]
    \includegraphics[width=0.5\textwidth]{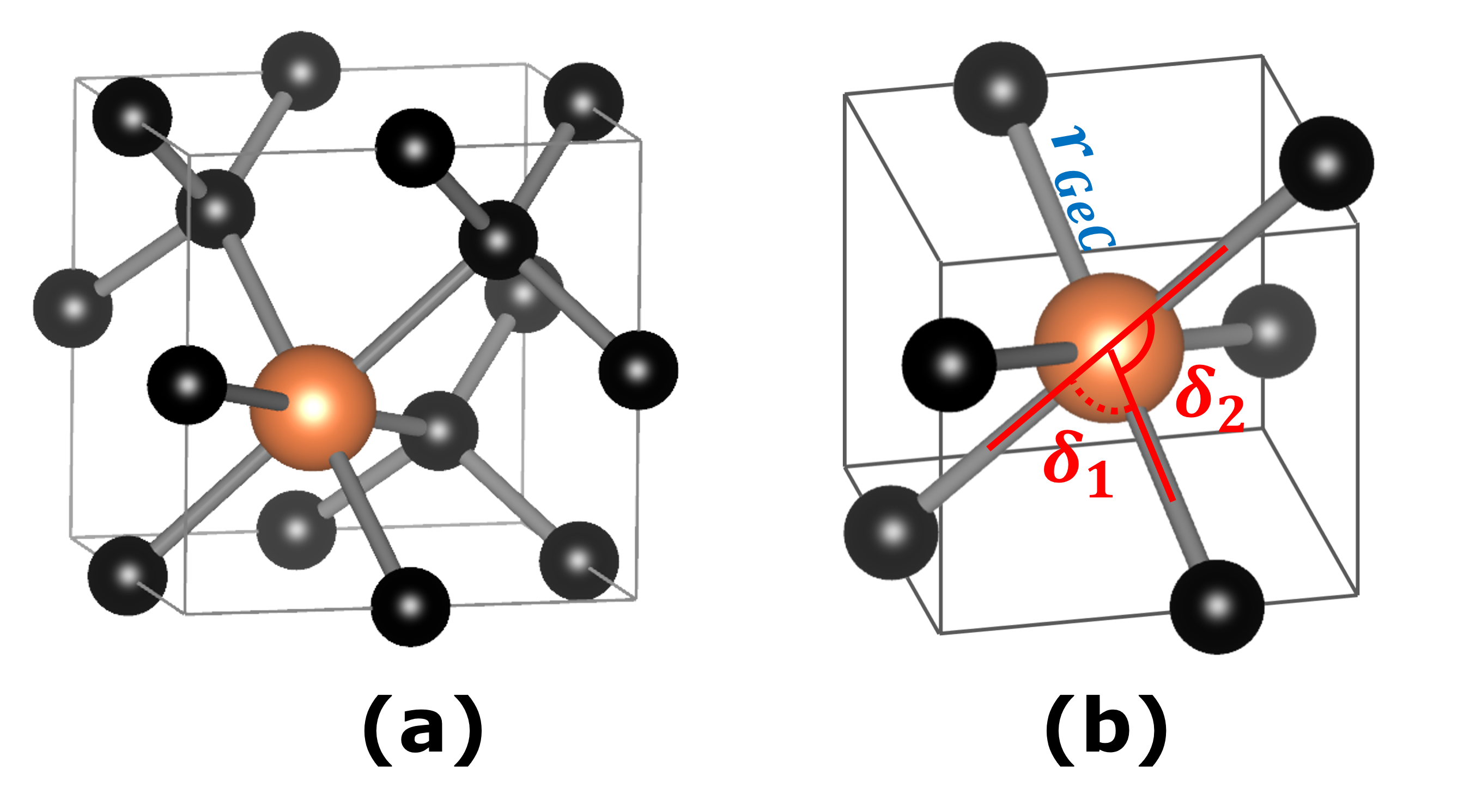}
    \caption{(a) Ball-and-stick representation of the GeV color center with split-vacancy structure in a $1\times 1\times 1$ conventional supercell. The Ge atom is indicated in orange. (b) Local atomic structure of the GeV color center, showing the six-coordinated Ge, and the six surrounding C atoms. The two different angles, $\delta_1$ and $\delta_2$, and bond length, $r_{GeC}$, are indicated.\label{Fig:GeV222structure}}
\end{figure}

\subsection{Color Center Model}\label{sec:model}
The GeV$^0$ color center in diamond is modelled using conventional supercells of four different sizes: $2\times 2\times 2$, $3\times 3\times 3$, $4\times 4\times 4$, and $5\times 5\times 5$. In each supercell, two neighboring C atoms are removed and replaced by a single Ge atom (\textit{cf.}, Fig.~\ref{Fig:GeV222structure}). Upon structure optimization, the Ge atom moves to a position corresponding to the center of the bond between the two former C atoms, which is why this defect is often referred to as a split-vacancy defect\cite{GossJP:2005PRB}. In its relaxed configuration, the Ge atom is six-coordinated and has a $D_{3d}$ symmetry. Inspection of the electron density shows the presence of six covalent bonds, highlighting the fact that the single Ge atom fills out all the available space. As Ge atoms are much larger than C atoms, the color center will locally strain the lattice.\\
In addition to local strain, two types of external strain are considered in this work as well: (i) hydrostatic strain and (ii) linear strain along the $\langle100 \rangle$ direction. Hydrostatic strain is modelled by modifying the system volume while retaining a cubic shape for the supercell (\textit{i.e.}, \textit{Cubic} optimization strategy). Five volumes are considered: $0, \pm1,$ and $ \pm2$ \% with regard to the PBE equilibrium volume. \review{The effective resulting strain for the different GeV$^0$ concentrations and functionals is obtained from the pressure-volume relation resulting from a third order isothermal Birch-Murnaghan equation-of-state (EOS) fit to the energy-volume data of the five points. Though the exact values vary slightly between the different systems, these five volumes span a pressure range of roughly $20$ GPa}. A first estimate of the equilibrium volume is obtained from the \textit{Skewed} optimization strategy (with both shape and volume relaxed), while the final equilibrium volume is defined as the minimum of a third order isothermal Birch-Murnaghan equation-of-state EOS fit\cite{BirchF:PhysRev1947, MurnaghanFD:PNAS1944}. For the linear strain along the $\langle 100\rangle$ direction, the equilibrium configuration is taken as a cubic cell with the PBE bulk diamond lattice parameter ($3.5704$ \AA). In addition, four strained configurations are generated by stretching the $\langle 100\rangle$ lattice parameter by $\pm 1$ and $\pm 2$ \%. The other two lattice parameters, however, are kept fixed (\textit{i.e.}, \textit{Bulk} optimization strategy). The resulting five energy-volume pairs are fitted using the same EOS, allowing for a comparable estimate of the experienced pressure.\\ 
\subsection{Experimental Details}\label{subsec:exp_details}
NCD films are grown using plasma-enhanced chemical vapor deposition (PECVD) on a \SI{5}{\milli\metre} $\times$ \SI{5}{\milli\metre} $\times$ \SI{0.5}{\milli\metre} (100)-oriented Ge substrate. Ge acts both as the substrate and as the solid dopant source simultaneously\cite{MaryJoy2024pp3873}. The substrates are cleaned by an exposure to hydrogen plasma in the ASTeX 6500 microwave (MW) PECVD system. 
To create a high nucleation density, the cleaned substrates are seeded with a water-based detonation nanodiamond colloid by drop-casting, followed by a deionised water rinse and spin-drying steps\cite{POBEDINSKAS2021148016}. The NCD film growth is carried out in the ASTeX 6500 MW PECVD system using a process gas mixture of \SI{99}{\percent} hydrogen and \SI{1}{\percent} methane with a total flow of \SI{400}{\sccm}. The working pressure and MW power are \SI{45}{\Torr} and \SI{3000}{\watt}, respectively. The growth temperature is \SI[separate-uncertainty]{720\pm20}{\celsius}, measured using a dualwavelength Williamson Pro92 pyrometer. With these process conditions, a deposition time of 1 h leads to an NCD film thickness of \SI{100}{\nano\metre}, monitored using the in-situ laser interferometry.\\ 
Room temperature photoluminescence (PL) measurements are carried out using Quantum Gem 532 laser with \SI{532}{\nano\metre} wavelength and \SI{1}{\milli\watt} power, focused via an objective with a numerical aperture of 0.95. For the GeV ZPL detection, a band-pass filter (ThorLabs FB600-10, center wavelength~=~\SI[separate-uncertainty]{600\pm2}{\nano\metre}, Full Width Half Maximum~=~\SI[separate-uncertainty]{10\pm2}{\nano\metre}) is used in the PL detection path for mapping.
%-------------------------------------
\section{Results and Discussion}
In the present work, three types of strain are considered: (a) local strain due to the presence of other defects, approximated via the color center concentration with the structures optimized using the \textit{Bulk}, \textit{Cubic} and \textit{Skewed} optimization strategies, (b) hydrostatic strain, using the \textit{Cubic} optimization strategy and (c) linear strain along the $\langle 100\rangle$ direction, using the \textit{Bulk} optimization strategy.

\begin{figure*}[!t]
    \includegraphics[width=\textwidth]{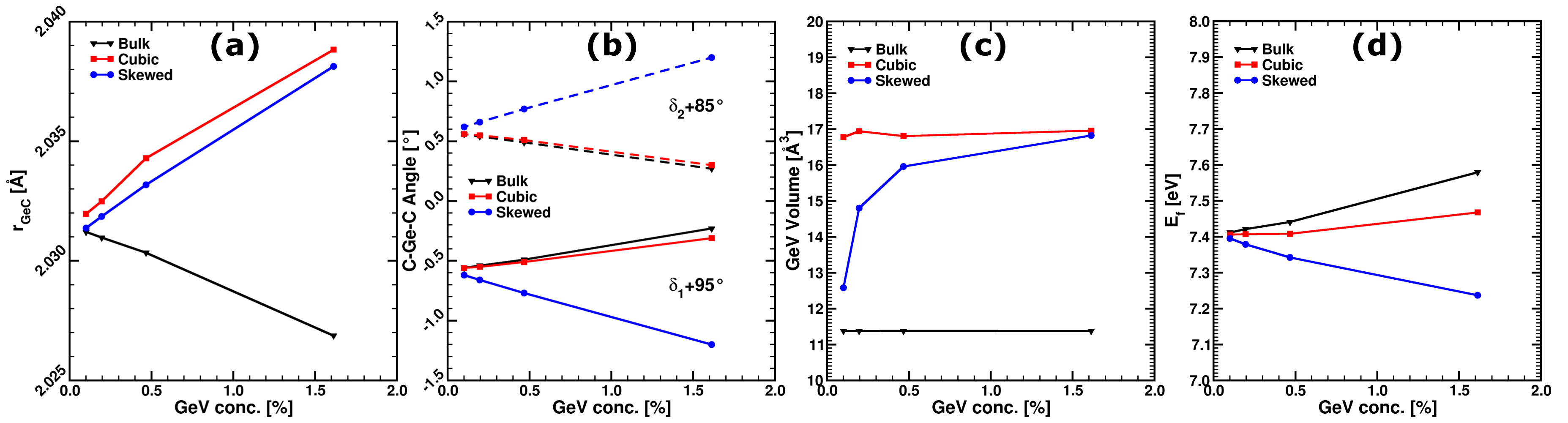}
    \caption{Structural properties of the GeV$^0$ color center, as a function of the concentration for the three different structure models: (a) the Ge-C bond length, $r_{GeC}$, (b) the complementary angles, $\delta_1$ and $\delta_2$, and (c) the color center volume. (d) PBE-level formation energy of the GeV$^0$ center as function of concentration. \label{Fig:StructureEvolution}}
\end{figure*}

\subsection{Color Center Concentration Effects}
The impact of the color center concentration is studied through the comparison of supercells of different sizes: the small $2\times2\times2$ supercell models a concentration of $1.61$ \%, whereas the larger $5\times5\times5$ models a concentration $0.10$ \%. Although this is much larger than the usual experimentally observed concentrations for GeV centers, which is of the order of ppm or even ppb\cite{BoldyrevKN:DiamRelatMater2022,iwasaki:2015SR,palyanov:2015sr}, the obtained results provide property trends that can be extrapolated towards the lower experimental concentrations. In addition, by considering three different sets of constraints during structure optimization (\textit{cf.}, Sec.~\ref{sec:model}) further insight is gained in local strain effects.\\
Irrespective of the applied constraints and color center concentration, the relaxed structure presented the expected $D_{3d}$ symmetry (\textit{cf.,} Fig.~\ref{Fig:GeV222structure}b)\cite{hepp:2014prl, GossJP:1996prl, GossJP:2005PRB, GossJP:2007prb, thiering:2018prx}. As a result, the Ge atom in the color center acts as an inversion center (\textit{cf.}, Table~S.I), which makes the system less responsive to external electric fields since there is no permanent dipole moment. There is a rise in overlap of the excited states with the optical ground state, ensuring an improved ZPL intensity\cite{RufM:2021JAppPhys}. From Figs.~\ref{Fig:StructureEvolution}a and b, it is clear that both the bond length and the angles in the color center ($r_{GeC}$, $\delta_1$ and $\delta_2$ respectively, \textit{cf.,} Fig.~\ref{Fig:GeV222structure}b) depend on the concentration, as well as the constraints employed during structure optimization. For both, the trends are linear with concentration, with the results converging toward the same value for low concentrations. The different observed trends can be understood as a consequence of the applied constraints. For the \textit{Bulk} systems, the lattice parameters are fixed on the PBE bulk lattice parameter, and as such, any structural relaxation involving a change in bond length is partially neutralized. It requires a sufficiently large supercell to have enough space for the C-C bonds between periodic copies of the color center to fully compensate for the distortion due to the Ge-C bonds \review{(\textit{i.e.} when the super cell is sufficiently large, the atoms furthest away from the GeV$^0$ center will have C-C bonds closer to those of bulk diamond.)}. The same can be said for the \textit{Cubic} systems, where the volume is optimized as well. However, in this case there is a slight overestimation due to the local distortion of the host lattice which requires a sufficiently large supercell to allow for C atoms in a non-distorted bulk environment. Extrapolating the Ge-C bond lengths to zero concentration gives $2.0315$ \AA\ for the \textit{Bulk} and \textit{Cubic} constrained systems, and $2.0309$ \AA\ for the \textit{Skewed} system. This is significantly larger than the calculated C-C bond length in bulk diamond of $1.546$ \AA, and slightly smaller than the Ge-C bond length of $2.146$ \AA\ in the \ce{GeC6H18} molecule. For the bond angles, it should be noted that the \textit{Skewed} system no longer presents orthogonal lattice vectors, but lattice vectors at angles of $90.64^{\circ}$ down to $90.04^{\circ}$ at color center concentrations of $1.61$ \% and $0.10$ \%, respectively. Extrapolation to zero concentration results in the same angle pair under all three constraints: $94.42^{\circ}$ and $85.58^{\circ}$.\\
The volume of the color center is calculated as the difference between the optimized system volume and the volume of the pristine diamond bulk supercell. For systems with constrained lattice parameters (\textit{Bulk}), the volume is by definition that of two C atoms. However, when the volume is relaxed as well, the color center volume increases to $16.9$ \AA$^3$, or $2.97$ times the volume of a single C atom. Note that for the \textit{Skewed} systems the volume strongly reduces for increasing cell size. This is an artefact resulting from the presence of Pulay stresses. Although this effect is known to play a significant role in weakly bonded flexible systems like Metal Organic Frameworks\cite{VanpouckeDannyEP:2015b_JPhysChemC}, the small effect in the diamond systems at hand--- $5-10$ meV in total energy for the $1000$ atom system---appears to be sufficient to artificially decrease the color center volume significantly. The \textit{Cubic} systems on the other hand are obtained through an EOS-fit which does not suffer Pulay stresses, and thus presents the correct color center volume. Furthermore, in the latter case, a quickly converging result is obtained.\\
In addition to a local deformation of the host lattice, the GeV$^0$ center also gives rise to a significant redistribution of the local electron density. To quantify this redistribution, the Hirshfeld-I atomic charges are calculated for the different systems. As expected, the variations with regard to the GeV$^0$ concentration are small, and mainly a consequence of the variations in the bond lengths. To define the charge of the color center (not to be confused with the charge state), we consider the Ge atom and the six nearest neighbor C atoms. Charge oscillations in subsequent shells of C atoms quickly decay toward zero atomic charges. The Ge atom is found to be positively charged $1.61$ electron, while all the nearest neighbor C atoms carry an identical negative charge of $-0.60$ electron, resulting in a total charge of the GeV$^0$ center $-2.0$ electron. These values are roughly independent of the GeV$^0$ concentration (\textit{cf.}, Fig.~S.7). The GeV$^0$ center appears to draw in significant additional charge from the host surrounding, an effect also seen in other point defects in diamond\cite{VanpouckeDEP:DiamRelatMater2019}. In comparison, we also calculate the Hirshfeld-I charges for the GeV$^+$ and GeV$^-$ systems, using charged $5\times5\times5$ supercells and manually set orbital occupancies. Total charges of $-1.6$ and $-2.7$ electron are found for GeV$^+$ and GeV$^-$, respectively. This shows the GeV always draws in electrons from its surroundings, making it negatively charged. This is not to be confused with the GeV being in the negative charge state, which is recognized by the occupation of the GeV $e_g$ states in the band gap.\\
The formation energy of the GeV$^0$ color center, as function of the concentration is presented in Fig.~\ref{Fig:StructureEvolution}d, and is calculated as
\begin{equation}
    E_f=E_{def}-N_C\mu_C-N_{Ge}\mu_{Ge},
\end{equation}
where $E_{def}$ represents the total (PBE) energy of the system with the GeV$^0$ center, $\mu_X$ the chemical potential and $N_X$ the number of atoms of the respective atomic species present. The bulk energy in diamond configuration for both Ge and C is chosen as chemical potential\review{, with the latter derived from a supercell of the same size as the supercell with the GeV$^0$ center, to have optimal compensation of errors}. The results are well converged for the largest supercells, as seen in Fig.~S.3, with the formation energy of the GeV$^0$ center being $\sim7.4$ eV for the three structure optimization strategies. Note that the differences between the three strategies show the structural impact of the color center to be rather long-ranged. The resulting strain energy can be considered to have a volumetric component and a shear component, leading to the differences between the \textit{Cubic} and \textit{Skewed} systems, with regard to the \textit{Bulk} systems, respectively.\\
Although GeV and NV centers arise from the substitution of two carbon atoms by another element, the reconstruction of the atomic lattice results in different point group representations. The GeV color center, with $D_{3d}$ symmetry has a higher symmetry than the NV with $C_{3v}$ symmetry (which is a subgroup of the former), the most important difference being the inversion center at the Ge position. As a result, they exhibit different electronic structures, which can be derived from group theory.

Ge has four valence electrons in a $4s^24p^2$ configuration, while the dangling bonds of the six neighboring C atoms contain one electron each, in one lobe of an $sp^3$ hybridized orbital (\textit{cf.}, Fig.~S.1). The energy levels associated with the GeV color center in diamond are thus often considered as molecular orbitals constructed as a Symmetry-Adapted Linear Combination of Atomic Orbitals (SALC or Symmetry-Adapted LCAO) from the Ge atom valence orbitals and the dangling bonds of the six adjacent C atoms, based on the symmetry of the orbitals. 
The three lowest C related orbitals combine with the Ge orbitals and form pairs of bonding and anti-bonding combinations, while the $e_{g}$ orbital is non-bonding. The molecular orbital diagram of the GeV$^0$ color center is shown in Fig.~S.1e.\\
In contrast to this simplified picture, practical DFT calculations provide not only these one-electron orbitals, but all one-electron orbitals for all valence electrons in the system. The calculated Kohn-Sham (KS) \review{energy levels}~at the $\Gamma$-point for the $2\times 2\times 2$- and $5\times 5\times 5$-supercells are presented in Fig.~S.2. As the KS orbitals associated with the GeV$^0$ color center are expected to be localized on the Ge atom and its six surrounding C atoms, they can be identified based on their site-projected wave function character, and the knowledge of the theoretically predicted energy levels. To validate these results, the KS \review{energy levels}~are also calculated for a minimal (passivated) cluster model of the color center: \ce{GeC6H18}. When C and Ge atoms are constrained in the geometry of the embedded color center a similar picture of the positions of the GeV$^0$ color center states arises (\textit{cf.}, Fig~S.2), showing the theoretically expected degeneracy and orbital symmetries. Note that if this molecular cluster would be allowed to relax, the symmetry would increase to the $O_h$ point group, in which a threefold degenerate state would be found below the highest occupied molecular orbital levels.\\
For the calculation of the ZPL, we consider excitations from the valence band maximum (VBM) to the empty $e_g$ states of the GeV$^0$ located in the bandgap of diamond. The choice for the VBM instead of the $e_u$ state is a consequence of the resonance between the VBM and the $e_u$ state, located just below the VBM\cite{GaliA:2013prb, thiering:2018prx, thiering:2023ptmpe}. \review{For comparison, we also calculated the ZPL positions using the $e_u$ state as reference, which we will refer to as ZPL$_{e_u}$ (\textit{cf.} Fig.~S.5)}. The ZPL is, by definition, a vertical excitation at the $\Gamma$-point of the first Brillouin zone, as no phonons carrying momentum are created.\\

\begin{figure}[t!]
    \includegraphics[width=0.4\textwidth]{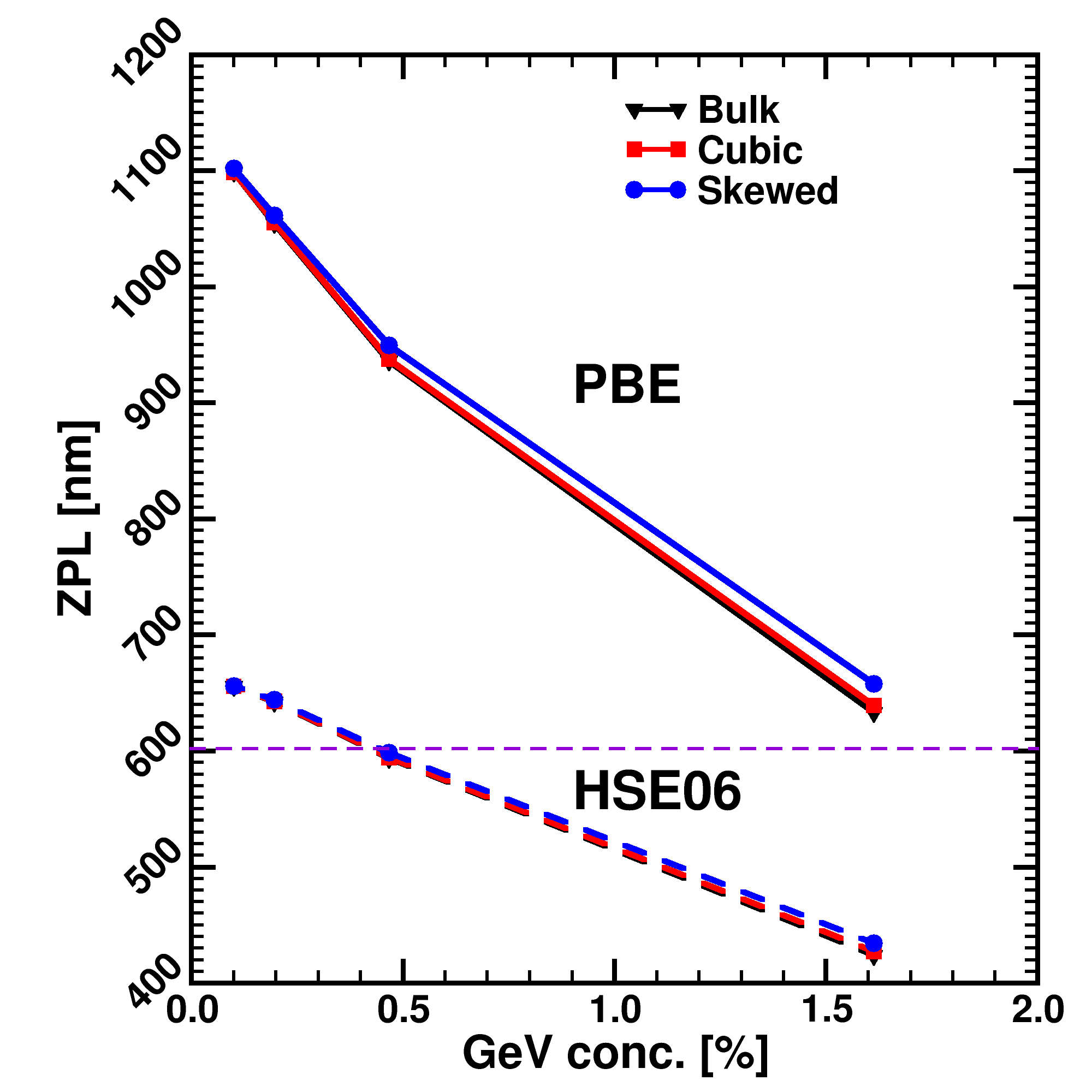}
        \caption{The calculated ZPL position, referenced to the VBM, as a function of the color center concentration for both PBE and HSE06 functionals. The dashed horizontal line indicates the position of the experimental value for the GeV$^-$.\label{fig:zpl}}
\end{figure}
The ZPL, calculated using the PBE and HSE06 functional, is shown in Fig.~\ref{fig:zpl}. The ZPL values are converged within $50$ meV for the largest supercell (\textit{cf.}, Fig.~S.3). As can be seen, both functionals present the same linear relation as function of the GeV$^0$ concentration. With the exception of the smallest supercell\cite{fn:c222shiftZPL}, all the HSE06 ZPL energies are shifted by $0.76$ eV, compared to the PBE values. This constant shift of the ZPL values between different functionals may be an indication that actual physical behavior is presented. The strong dependence of the absolute position of the ZPL on the selected functional, however, causes this to be a difficult experimental parameter to predict theoretically. \review{One should thus be careful with assignment based on calculations at a single concentration, for example, the HSE06 ZPL position from the $3\times3\times3$ supercell fits the experimental ZPL position of the GeV$^-$ rather nicely}. Despite this limitation, the known quality relation between the functionals is also present\cite{PerdewJacobs:AIPConfProc2001}. The HSE06 results are shifted to lower wavelengths, compared to PBE, closer to the experimentally observed value of \SI{602}{\nano\metre} for GeV$^-$. The overestimation of the ZPL position by the PBE functional is not unexpected, as it is well known that PBE (and local functionals in general) underestimate the bandgap as well as the position of states within the bandgap (\textit{cf.}, Fig.~S.4).\\ 
Although the concentrations considered are much higher than the experimental values (ppm range), the calculated shift of the ZPL as function of the GeV$^0$ concentration is an indication that the proximity of GeV$^0$ centers may give rise to a shift (or peak-broadening) in the spectrum of an ensemble of GeV$^0$ centers.\\  
\review{As a final remark, let us briefly consider the impact of our choice of defining the ZPL as an excitation from the VBM. If one used the $e_u$ state instead, a qualitatively similar picture arises with the ZPL shifted to lower wavelength (or higher energies). The resulting ZPL$_{e_u}$ are shown in Fig.~S.5. In addition, for low concentrations the trends become non-linear. This is a consequence of the non-linear shift of the position of the $e_u$ state with regard to the VBM as function of the color center concentration (\textit{cf.}, Fig.~S.5c). In two additional calculations (PBE level, \textit{Bulk} systems) using a $6\times6\times6$ and $7\times7\times7$ conventional supercell ($0.06$ and $0.04$ \% GeV$^0$) the $e_u$ state moves to a mere $-299$ and $-223$ meV below the VBM, from the $-415$ meV at $0.1$  \% GeV$^0$. In contrast, for the $215$ atom supercell ($0.47$ \% GeV$^0$) a value of $-723$ meV is found, in excellent agreement with the value of $-0.71$ calculated by Iwasaki and coworkers for the same system size and functional \cite{iwasaki:2015SR}. Also Goss and coworkers present a similar position of the $e_u$ state using a $215$ atom supercell and LSDA functional\cite{GossJP:2005PRB}. From these results, and the agreement with existing work, it appears that the $e_u$ state shifts to the VBM for the GeV$^0$ center in the dilute limit}.

\begin{figure*}[t!]
    \includegraphics[width=\textwidth]{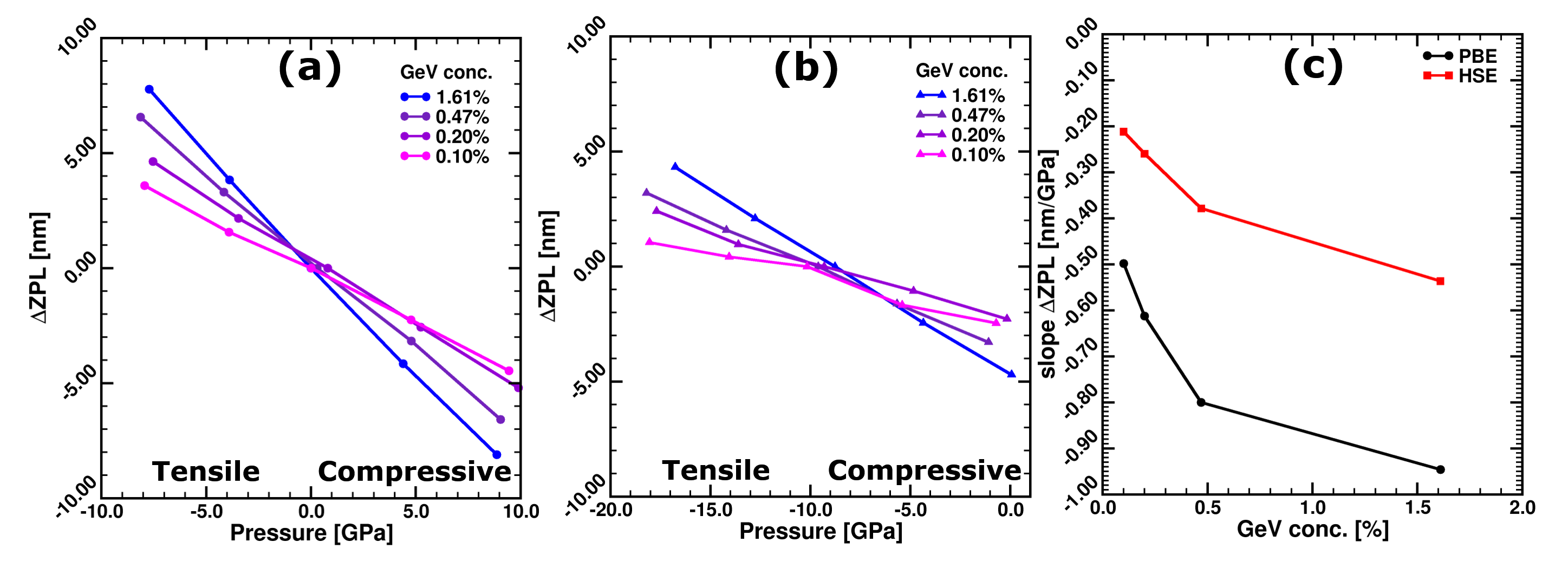}
        \caption{ZPL shift ($\Delta$ZPL) under hydrostatic stress as a function of pressure, for different color center concentrations. Calculations are performed using the PBE (a) and HSE06 (b) functionals. The off-set from zero-pressure equilibrium for the HSE06 calculations is a consequence of the use of PBE optimized geometries\cite{fn:HSE_ZPLshift}. The pressure is calculated from the EOS, with a positive and negative sign corresponding to compressed and expended systems, respectively. For all calculations the middle point represents the equilibrium configuration. Points experiencing compressive or tensile stress are indicated for convenience. (c) The concentration dependent slope of the ZPL shifts (pressure coefficient) presented in (a) and (b).\label{fig:hydro}}
\end{figure*}
\subsection{Hydrostatic Strain}\label{sec:hydro}
To compare the effects of the strain at different GeV$^0$ concentrations, the relative shift of the ZPL position, $\Delta$ZPL, (shown in Fig.~\ref{fig:hydro}) is chosen as indicator. The indicated pressure is obtained from the fitted third order Birch Murnaghan EOS, and offsets from zero for the equilibrium configuration are a consequence of the used methodology\cite{fn:HSE_ZPLshift}. It is important to note that under hydrostatic strain, the symmetry is not altered, and as such the degeneracy of the $e_g$ and $e_u$ states is retained with a numerical accuracy of $10^{-3}$ meV. This means no peak splitting or broadening is expected.\\
For all concentrations, and both functionals, a clear linear relation between $\Delta$ZPL and the experienced pressure is found. As the slope (\textit{i.e.} pressure coefficient) is the same for compressive and tensile strain, a red-or blue-shift is expected, respectively. The pressure coefficient (or slope of the $\Delta$ZPL--pressure relation) depends on the GeV$^0$ concentration as well as the functional (\textit{cf.}, Fig.~\ref{fig:hydro}c). Note, however, that the different functionals do show the same qualitative picture. Based on the results of the larger supercells, the pressure coefficient can be extrapolated (\textit{cf.}, Fig.~\ref{fig:hydro}c) to experimentally relevant concentrations, and is found to have a magnitude of $0.38$ and $0.15$ nm/GPa (or $0.32$ and $0.38$ meV/GPa, \textit{cf.} Table~S.II) for PBE and HSE06, respectively. \review{For similar work on the SiV$^-$ color center, Lindner and coworkers\cite{lindner:2018njp} calculated, using a $512$ atom supercell and the PBE functional, a pressure coefficient of the order of $1$ nm/GPa is obtained for various stresses, in agreement with experimental ZPL shifts of $0.52$ up to $4$ nm/GPa. Specifically for the hydrostatic situation, the calculated pressure coefficient is of the order of $0.5$ nm/GPa for stresses below $5.5$ GPa, indicating the obtained values for GeV$^0$ could be reasonable}.\\ 
\review{In the case that the $e_u$ state is used as reference, the absolute values of the pressure coefficients increase by a factor of $2-3$(meV/GPa) or $1.5-2$(nm/GPa) (\textit{cf.} Table~S.II), but are expected to coincide in the dilute limit when $e_u\rightarrow$VBM.\\
Krivobok \textit{et al.}\cite{KrivobokV:2020PRB} attributed a ZPL line at $1.979$ eV ($626$ nm) to the GeV$^0$ center in contrast to earlier work of the same group attributing it to the vibronic sideband of GeV$^-$. For this ZPL a pressure coefficient of $2.9$ meV/GPa was measured, almost the same as for GeV$^-$. This value is higher than the calculated dilute limit value above, though it is well within the range of values calculated for the mid to high concentration cells. In comparison, Vindolet \textit{et al.}\cite{VindoletB-2022APS} reported an experimental blue-shift of $2.7$ meV/GPa ($\sim 0.8$ nm/GPa\cite{fn:VindoletGuess}) under a hydrostatic pressure in the range of $20$ to $40$ GPa range for the GeV$^-$ center. In the same work, these authors present a calculated value of $2.90$ meV/GPa using the SCAN functional and a $512$ atom supercell ($0.2$\% GeV$^-$), while Ekimov \textit{et al.}\cite{ekimovE:2019jetp} calculated a value of $3.1$ meV/GPa for GeV$^-$ using the HSE06 functional and non-conventional $88$ and $233$ atom supercells, $1.1$ and $0.4$ \% GeV$^-$, respectively, also indicating a shift to smaller values with increasing cell size}. These calculations, at single concentration points, were in agreement with the experimentally obtained values for the GeV$^-$ center, indicating the charge state to have a non-negligible influence.\\
The origin of the concentration dependence of the pressure coefficient (\textit{cf.,}~Fig.~\ref{fig:hydro}c) could be understood from a simple particle in a box model: the splitting between energy levels depends on the size of the box, with smaller boxes resulting in larger splittings. In the current case, the supercells are modified using the same relative change in lattice parameters, irrespective of their corresponding GeV$^0$ concentrations. However, as we noted in the previous section, the local deformation of the host lattice due to the presence of the GeV$^0$ center is long-ranged, and decays linearly with the GeV$^0$ concentration. As such, within the particle in a box approximation of the GeV$^0$ center, the change of the associated box size due to a relative change in the supercell size, will become less pronounced for the larger supercells as there is more room for both deformations to be accommodated. Therefore, lower concentration supercells also experience a smaller ZPL shift at the same pressure.
\begin{figure*}[t!]
    \includegraphics[width=\textwidth]{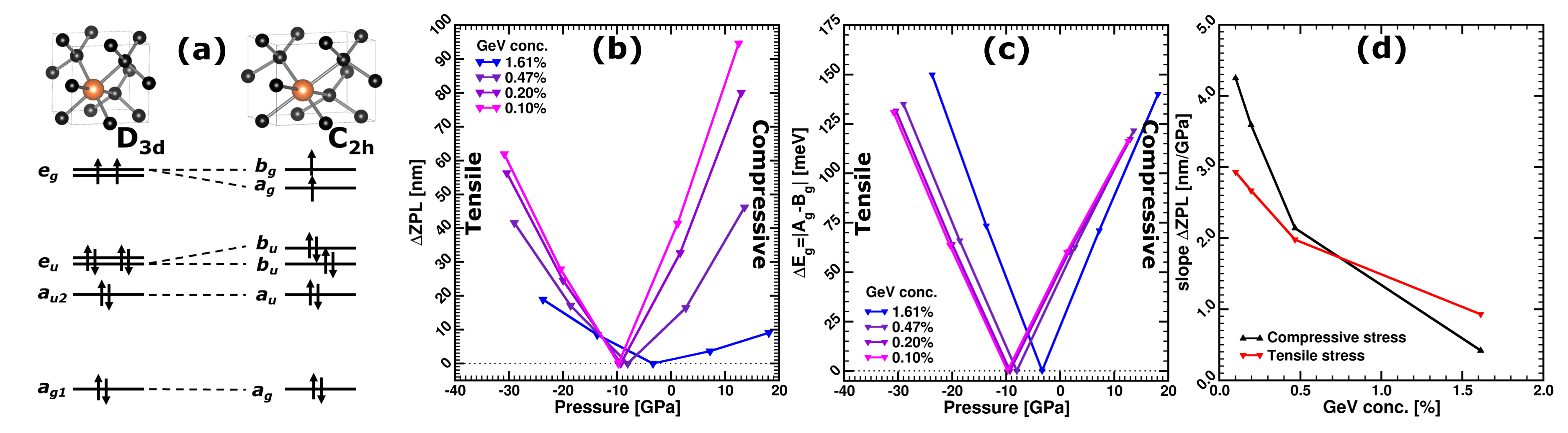}
    \caption{(a) Splitting of the GeV$^0$ energy levels due to linear strain. Left: Theoretical energy levels for the GeV$^0$ center in the $D_{3d}$ symmetry. Right: Theoretical energy levels for the GeV$^0$ center with linear strain applied along the $\langle100\rangle$ direction, resulting in a $C_{2h}$ symmetry. The initially degenerate energy levels in the $D_{3d}$ symmetry, \textit{i.e.} $e_g$ and $e_u$, are split into $a_g$, $b_g$, $a_u$ and $b_u$. The calculated $\Delta$ZPL (b) and level splitting (c) under linear train, using the HSE06 functional, for different GeV$^0$ concentrations\cite{fn:HSE_ZPLshift}. (d) The pressure coefficient (\textit{i.e.}, slope of $\Delta$ZPL) as function of the GeV$^0$ center concentration for both compressive and tensile stress, calculated using the HSE06 functional.\label{fig:linstrain}}
\end{figure*}

\subsection{Linear Strain along the $\langle100\rangle$ Direction}\label{sec:lin}
Similar to the hydrostatic strain, the relative shift of the ZPL position, $\Delta$ZPL, is calculated under linear strain. This to allow for comparison between different GeV$^0$ concentrations (\textit{cf.}, Fig.~\ref{fig:linstrain}). The indicated pressures are the result of third order Birch Murnaghan EOS fits to the Energy--Volume data resulting from the different linearly strained models\cite{fn:HSE_ZPLshift}. In contrast to the hydrostatic case, the symmetry of the system is broken and reduced to a $C_{2h}$ symmetry. As a result the degeneracy of the $e_g$ states is lifted, resulting in $b_g$ and $a_g$ states (\textit{cf.}, Fig.~\ref{fig:linstrain}a).\\
The shift of the ZPL due to tensile or compressive strains for the different GeV$^0$ concentrations is presented in Fig.~\ref{fig:linstrain}b for the HSE06 functional (results for PBE are given in SI). As for the hydrostatic case, the shift is more pronounced when the PBE functional is used, though the qualitative behavior is the same for both functionals, with the HSE06 results expected to be quantitatively more accurate. An important difference with the hydrostatic strain is that whereas the former converged to lower $\Delta$ZPL values for lower GeV$^0$ concentrations, in the case of linear strain, the convergence is toward a larger value. A second qualitative difference with the hydrostatic case is found in the fact that both tensile and compressive strain give rise to a red-shift. It is interesting to note that in the case of linear strain, the calculated pressure coefficient differs for tensile and compressive strain (\textit{cf.}, Fig.~\ref{fig:linstrain}d). Extrapolation to experimentally relevant GeV$^0$ concentrations results in a red-shift of $3.2$ or $4.8$ nm/GPa ($8.1$ or $11.7$ meV/GPa), for tensile and compressive stress, respectively. This is more than one order of magnitude stronger than for the hydrostatic case. This is in line with the results for the SiV$^-$ color center, where also a larger pressure coefficient was calculated in the case of a linear strain along $\langle100\rangle$ compared to hydrostatic strain\cite{lindner:2018njp}.\\
The theoretically expected splitting of the $e_g$ state into the $b_g$ and $a_g$ states is also obtained from the DFT calculations (\textit{cf.}, Fig.~\ref{fig:linstrain}c and S.6b), and shows a comparable scale for PBE and HSE06. For both cases, the splitting behaves linear with the applied stress, with the results for HSE06(PBE) converging to $6.2$($6.0$) and $5.2$($5.2$) meV/GPa for tensile and compressive stress, respectively. Combined with the shift of the ZPL postion, this splitting is expected to give rise to a deformation and peak broadening of the experimentally observed ZPL line.\\

\subsection{Experimental Observations and Theoretical Interpretation}

\begin{figure*}[tb!]
    \includegraphics[width=\textwidth]{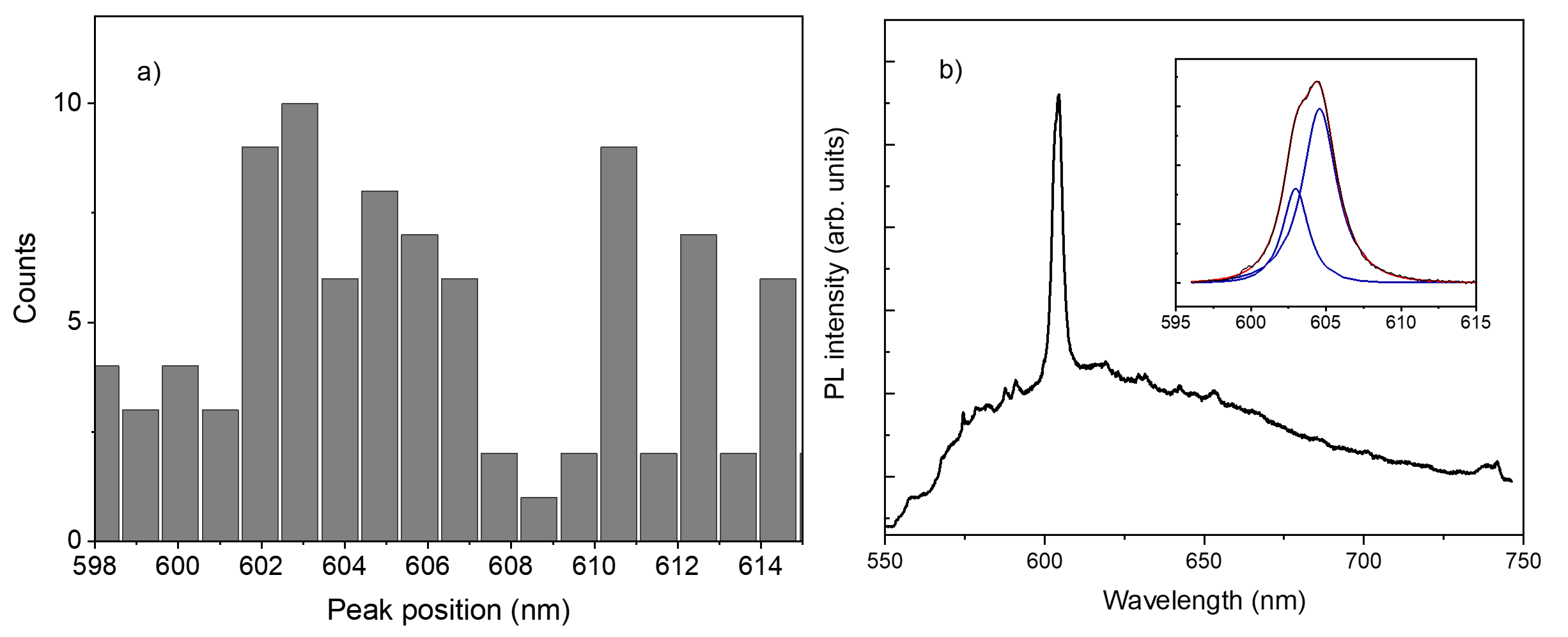}
    \caption{(a) Distribution of the PL positions and (b) an example of a room temperature PL spectrum of the NCD on Ge substrate. The inset shows a zoom-in of the GeV ZPL with two fitted peaks. Reused with permission from \cite{MaryJoy2024pp3873}. \label{Fig:GeV_histogram}}
\end{figure*}
Experimental studies carried out on NCD growth on Ge substrates have reported successful formation of GeV centers in diamond films\cite{MaryJoy2024pp3873, RalchenkoVG:2015BullLebPhysInst}. In our previous study, we reported the variation of the GeV ZPL position and attributed it to the residual stress in the NCD film (\textit{cf.},~Fig.~\ref{Fig:GeV_histogram}a)\cite{MaryJoy2024pp3873}. The remaining peaks observed in the spectrum (\textit{e.g.},~Fig.~\ref{Fig:GeV_histogram}b) may be surface-related or defects associated with nitrogen, but their exact origins have yet to be determined. This, however, falls outside of the scope of the current work.\\
Before drawing any comparisons with the DFT modelling, there are several important experimental observations to highlight when discussing the PL spectra of GeV centers. 
% First experimental result/factor to consider
First, since Ge has a higher thermal expansion coefficient than diamond, a NCD film grown on a Ge substrate will experience a compressive residual stress once the sample is cooled down after high temperature diamond deposition. This compressive stress comes along with a tensile stress perpendicular to the substrate surface. For the growth conditions discussed in the experimental details, the in-plane residual compressive stress was estimated to be \SI{0.85\pm0.15}{\giga\pascal} in magnitude and the out-of-plane tensile stress was estimated to be \SI{0.21\pm0.03}{\giga\pascal} in magnitude\cite{MaryJoy2024pp3873}. 
% Second experimental result/factor to consider
Second, SEM images of the NCD film (\SI{100}{\nano\metre} thick) grown on a Ge substrate show that the diamond grains are randomly oriented\cite{MaryJoy2024pp3873}. Since the GeV center in diamond is always oriented along the $\langle111\rangle$ direction, the apparent random orientation of the diamond grains in the NCD film will result in equally apparent randomly oriented GeV centers. 
% Third experimental result/factor to consider
Third, the analysis of our previous work (\textit{cf.} Fig.~\ref{Fig:GeV_histogram}a), revealed the presence of both red- and blue-shifted ZPL (although with more observations of red-shifts than blue-shifts, the amplitude of the former being larger than the latter). 

One example of a GeV spectrum is shown in Fig.~\ref{Fig:GeV_histogram}b, with the most significant PL peak observed at about \SI{604}{\nano\metre}. As is visible in the inset, the ZPL line has a clear shoulder and is formed by two contributions. \review{The GeV ZPL is fitted to a Voigt function using Origin. The center wavelength position of the fitted peaks is located around \SI{603\pm0.1}{\nano\metre} and \SI{604.5\pm0.1}{\nano\metre}, which corresponds to an energy separation of about $5.1$ meV}. Considering the results of the linear strain model---theoretically calculated ZPL shift of about $3.2$ nm/GPa and $4.8$ nm/GPa ($8.1$ or $11.7$ meV/GPa) for tensile and compressive stress, respectively---the observed ZPL shift of about \SI{1}{\nano\metre} with regard to a non-strained GeV center corresponds to an experienced stress of less than $1$ GPa. Considering the observed splitting, which also is part of the linear strain model, the observed $5.1$ meV is in line with an experienced stress of $1$ GPa or less. In addition, the larger splitting of the peak by about \SI{1.5}{\nano\metre} is in qualitative agreement with the calculated HSE06 values for the linearly strained system, which also present a larger effect due to the splitting than the shift.\\

The spectrum presented in Fig.~\ref{Fig:GeV_histogram}b can thus be understood as the result of a linearly strained system. However, as we discussed in the previous section, a linearly strained system gives rise to red-shifted ZPL for both compressive and tensile strain. In contrast, in Fig.~\ref{Fig:GeV_histogram}a both red-shifted and blue-shifted ZPL may be present.
Looking back at the two external strain models discussed in this work, the vector along the GeV orientation behaves qualitatively different under hydrostatic and linear strain. In the case of hydrostatic strain, this vector will change in magnitude but not in orientation. In contrast, for the linear strain, this vector will change in orientation (breaking the symmetry). The in-plane aligned GeV centers experience an isotropic compressive stress (in the plane). This could be similar to the hydrostatic stress modelled above (depending on the orientation of the surrounding C atoms, the direction of the GeV orientation will only change in magnitude) and for compressive stress is predicted to give rise to blue-shifted ZPL. In contrast, note that for a linear compressive stress, a red-shifted ZPL is predicted. As such, this would clarify the observed blue-shifted ZPL in Fig.~\ref{Fig:GeV_histogram}a.
%% a) pure perpendicular would be 111 stress which retains symmetry so red and blue shift
%% b) not perpendicular can be 100 and 110 stress which gives C2h only red-shift.
For an out-of-plane oriented GeV center an anisotropic strain will be experienced resulting in a breaking of the symmetry and a change in the orientation of the GeV (compared to the non-strained case). As such, a situation similar to the linearly strained GeV$^0$ center is observed. Depending on the exact orientation of the GeV and the ratio of the in-plane compressive and perpendicular tensile strain, both the compressive and tensile picture of the linear strain model come into play. However, as we noted, both give rise to red-shifted ZPL, be it with a different magnitude. Furthermore, the magnitude of these ZPL-shifts are shown to be much larger in the linear strain case than the hydrostatic strain case.\\
Considering Fig.~\ref{Fig:GeV_histogram}a in detail, one notes there are more red-shifted cases than blue-shifted cases. This is in line with the expectation of a distribution of the GeV orientations. The exact nature of this distribution will influence the details of the observed ZPL shift distribution, but goes beyond the scope of the current work. Finally, the red-shifted ZPL go to larger values than the blue-shifted cases, with a bimodal distribution for the former, which is in line with the expectations from the linear strain model showing a difference between tensile and compressive stress, while showing a stronger shift than hydrostatic strain.\\

\section{Conclusion}
The impact of strain on the ZPL of the GeV$^0$ color center in diamond is investigated using DFT, and compared to experimental observations. Three sources of local strain are considered: the proximity of other color centers, externally applied hydrostatic strain, and externally applied linear strain along the $\langle100\rangle$ direction. For both the PBE and HSE06 functional the same qualitative linear relations between the ZPL position and color center concentration, as well as between the ZPL shift and the applied external strain are obtained. The former allows for extrapolation to experimentally relevant concentrations, though some deviation from the experimental value remains to be expected due to the choice of the functional, with higher rung functionals expected to approximate the experimental reality better. The observation of a concentration dependence of the ZPL position indicates one should be careful when proposing a direct link between the calculated ZPL position for one specific concentration and the values measured at ppm or ppb levels.\\
In the case of an externally applied strain, qualitatively different behavior is observed for different types of strain. Hydrostatic strain retains the color center symmetry, with the shift of the ZPL decreasing with decreasing GeV$^0$ center concentration. Extrapolation of the obtained linear trend suggests both red- and blue-shifted ZPL are expected for tensile and compressive strain, respectively. The size of the ZPL shift is the same for tensile and compressive stress, and extrapolation to experimental color center concentrations based on the calculated values is about $0.15$($0.38$) nm/GPa or $0.32$($0.38$) meV/GPa for HSE06(PBE) functional. In the case of linear strain along the $\langle100\rangle$ direction, only red-shifted ZPL are expected. The extrapolated red-shift for experimentally relevant concentrations is an order of magnitude larger for the HSE06 calculations, namely about $3.2$ and $4.8$ nm/GPa (or $8.1$ and $11.7$ meV/GPa) for tensile and compressive stress, respectively. In addition to this much larger shift, the degeneracy of the $e_g$ GeV$^0$ state is lifted, giving rise to a splitting between the two levels with a size of about $6$ meV/GPa. Comparison to an experimentally observed GeV spectrum presents a qualitative agreement, supporting a symmetry breaking strain and showing a complex picture of significant variation of the strains experienced by individual GeV centers in NCD. Despite the inherent complexity present in the experimental sample and measurement, as well as the limitations present in first principles calculations, a qualitative understanding of the experimental observations can be obtained, providing a guide for future work.

\section*{Acknowledgement}
\indent DEPV thanks Vadim Sedov for the fruitful discussions on the subject of defects in diamond. This work was financially supported by the Special Research Fund (BOF) via the BOF project R14181, the Methusalem NANO network, interuniversity BOF (BOF 23IU12), Research Foundation Flanders (FWO) projects G0D1721N and G0A0520N. The computational resources and services used in this work were provided by the VSC (Flemish Supercomputer Center), funded by the FWO and the Flemish Government--department EWI.

%\section*{Data Availability}
%The data that support the findings of this study are available from the corresponding author upon reasonable request.

%******************************************************
% BIBLIOGRAPHY
%*******************************************************
% Put in \nocite{*} so all entries in the bibliography are included
%\nocite{*}
% This GATHER command is useful for when you want to use WinEdt's Gather functionality, i.e., type
% \cite{} and a popup box appears with all of your citations to choose from.  Leave the % on the next line.
% The commented way of writing Gather is the only correct way of doing it !!!
%%GATHER{GeV.bib}
%%GATHER{GeV_review.bib}
%%GATHER{notes.bib}
%\bibliography{GeV,GeV_review,notes}

\begin{thebibliography}{75}
\bibitem [{fn:({\natexlab{a}})}]{fn:c222shiftZPL}%
  \BibitemOpen
  %\href@noop {} {} ({\natexlab{a}}),\ 
  \bibinfo {note} {{F}or the smallest super cell, this shift was $0.97$ eV for the three optimisation strategies.}\BibitemShut {Stop}%
\bibitem [{fn:({\natexlab{c}})}]{fn:VindoletGuess}%
  \BibitemOpen
  %\href@noop {} {} ({\natexlab{c}}),\ 
  \bibinfo {note} {{T}his value was estimated based on the $8$ and $50$ GPa curves, presented in figure 2 of same paper: $\sim 35$ nm /$42$ GPa = $\sim 0.8$ nm/GPa.}\BibitemShut {Stop}%
\end{thebibliography}

%apsrev4-2.bst 2019-01-14 (MD) hand-edited version of apsrev4-1.bst
%Control: key (0)
%Control: author (8) initials jnrlst
%Control: editor formatted (1) identically to author
%Control: production of article title (0) allowed
%Control: page (0) single
%Control: year (1) truncated
%Control: production of eprint (0) enabled
%

\end{document}